\documentclass[a4paper,fleqn,usenatbib,useAMS]{mnras}

\usepackage{caption}
\usepackage{subcaption}
\usepackage{graphicx}	% Including figure files
\usepackage{amsmath}	% Advanced maths commands
\usepackage{amssymb}	% Extra maths symbols
\usepackage{multicol}        % Multi-column entries in tables
\usepackage{bm}		% Bold maths symbols, including upright Greek
\usepackage{pdflscape}	% Landscape pages
\usepackage{hyperref}
\usepackage{xcolor}
\usepackage{lipsum}

\usepackage[T1]{fontenc}
\usepackage{ae,aecompl}

\usepackage{newtxtext,newtxmath}
\title[Instability of Saturn's Retrograde Co-orbitals]{On the Instability of Saturn's Hypothetical Retrograde Co-orbitals}

\author[Yukun Huang et al.]{
Yukun Huang,$^{1}$\thanks{E-mail: puluobi@gmail.com}
Miao Li,$^{1}$
Junfeng Li$^{1}$
and Shengping Gong$^{1}$\thanks{E-mail: gongsp@tsinghua.edu.cn}
\\
$^{1}$School of Aerospace Engineering, Tsinghua University Beijing, China, 100086\\
}

\date{Accepted XXX.\@ Received YYY.}

\pubyear{2018}

\begin{document}\label{firstpage}
\pagerange{\pageref{firstpage}--\pageref{lastpage}}
\maketitle

\begin{abstract}
	We find an interesting fact that fictitious retrograde co-orbitals of Saturn, or small bodies inside the retrograde 1:1 resonance with Saturn, are highly unstable in our numerical simulations. It is shown that in the presence of Jupiter, the retrograde co-orbitals will get ejected from Saturn's co-orbital space within a timescale of 10 Myr. This scenario reminds us of the instability of Saturn Trojans caused by both the Great Inequality and the secular resonances. Therefore, we carry out in-depth inspections on both mechanisms and prove that the retrograde resonance overlap, raised by Great Inequality, cannot serve as an explanation for the instability of retrograde co-orbitals, due to the weakness of the retrograde 2:5 resonance with Jupiter at a low eccentricity. However, we discover that both $\nu_5$ and $\nu_6$ secular resonances contribute to the slow growth of the eccentricity, therefore, are possibly the primary causes of the instability inside Saturn's retrograde co-orbital space.
\end{abstract}


%%%%%%%%%%%%%%%%%%%%%%%%%%%%%%%%%%%%%%%%%%%%%%%%%%

%%%%%%%%%%%%%%%%% BODY OF PAPER %%%%%%%%%%%%%%%%%%

\section{Introduction}\label{sec:1}
The absence of Saturn Trojans has stirred interests of astronomers for decades. With the development of the astronomical observation technology, Trojans of Uranus \citep{Alexandersen:2013dq}, Neptune \citep{Marzari:2003eh} have been discovered in recent years. However, until now, not a single Saturn Trojan has been found in any surveys. Considering the abundance of Jupiter Trojans, it is so aberrant for the second biggest planet in the Solar System to have no neighbours around its Lagrange points. After realizing the anomaly of Saturn, astronomers quickly spotted Jupiter as the culprit through numerical integrations \citep{Innanen:1989dv}. Further numerical surveys by \citet{Holman:1993iw} demonstrated that there are two holes near the triangular Lagrange points of Saturn, implying that Trojans with small amplitude are highly unstable.

Aiming at explaining the instability in Saturn's co-orbital region, \citet{delaBarre:1996he} analysed two mechanisms that may lead to this phenomenon, the Great Inequality and the $\nu_6$ secular resonance, with both the numerical method and the Hamiltonian perturbation theory. \citet{Marzari:2000eq} did more investigation on the secular dynamics and identified that the mixed $2\varpi_S - \varpi_J - \varpi_T$ resonance also accounted for the eccentricity growth of Saturn Trojans. Regarding the explanation of Great Inequality, \citet{Nesvorny:2002il} demonstrated that the instability was duo to the overlap between two mean motion resonances. With a planar bi-circular model, it is shown that chaos can be generated for orbits that are close to the tadpole centers with $e > 0.13$. Therefore, the view of both Great Inequality and secular resonances contributing to the instability and clearing out any potential Saturn Trojans has been gradually formed. Recently, \citet{Hou:2013ic} re-examined the stability problem in the synodic frame with the aid of the frequency analysis, which again, corroborated the combined roles of the resonance overlap and the secular resonance.

It seems that former researchers have put an end to the problem. However, the latest discoveries of the first retrograde co-orbital body of Jupiter \citep{Wiegert:2017fj} and potential retrograde co-orbitals of Saturn \citep{Li:2018kn} have once again ignited our curiosity to the co-orbital region of Saturn. In the process of studying the dynamics of retrograde resonances \citep{Huang:2018ey,Huang:2018cz}, we noticed that Saturn's retrograde co-orbital region is significantly unstable. The fact that retrograde co-orbitals are nowhere to be found in our numerical simulations urges us to reinspect Saturn's co-orbital space for retrograde orbits.

In this paper, we first reported the instability emerged from the retrograde co-orbital region of Saturn in terms of a wide range of eccentricities and inclinations. Afterwards, through a comparative simulation, we confirmed that it is the gravitational influence of Jupiter that clears out potential retrograde co-orbitals of Saturn. Then, in Sec.~\ref{sec:3}, we analyse both mechanisms of Great Inequality and secular resonances and rule out the possibility that resonance overlap may play a part. In the end, we verify that both $\nu_5$ and $\nu_6$ resonances have a major impact on destabilizing co-orbital bodies, corroborating that secular resonances are the primary factors leading to the vanishing of Saturn's hypothetical retrograde co-orbitals.

\begin{figure*}
	\centering
	\begin{subfigure}{0.33\textwidth}
        \centering
		\includegraphics[width=\textwidth]{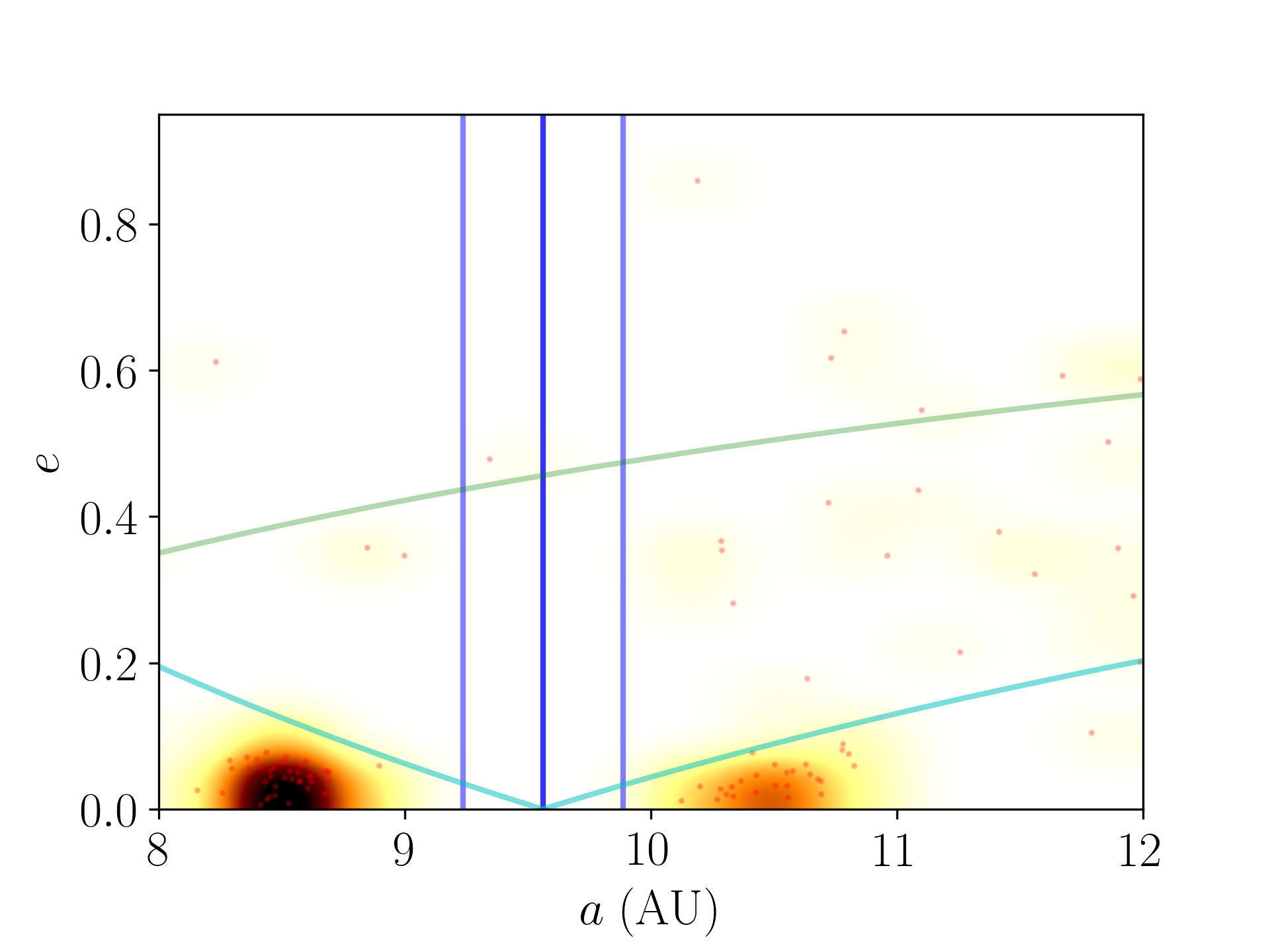}
        \caption{Planar case ($i = 180\degr$)}\label{fig:planar}
    \end{subfigure}
	\begin{subfigure}{0.33\textwidth}
        \centering
		\includegraphics[width=\textwidth]{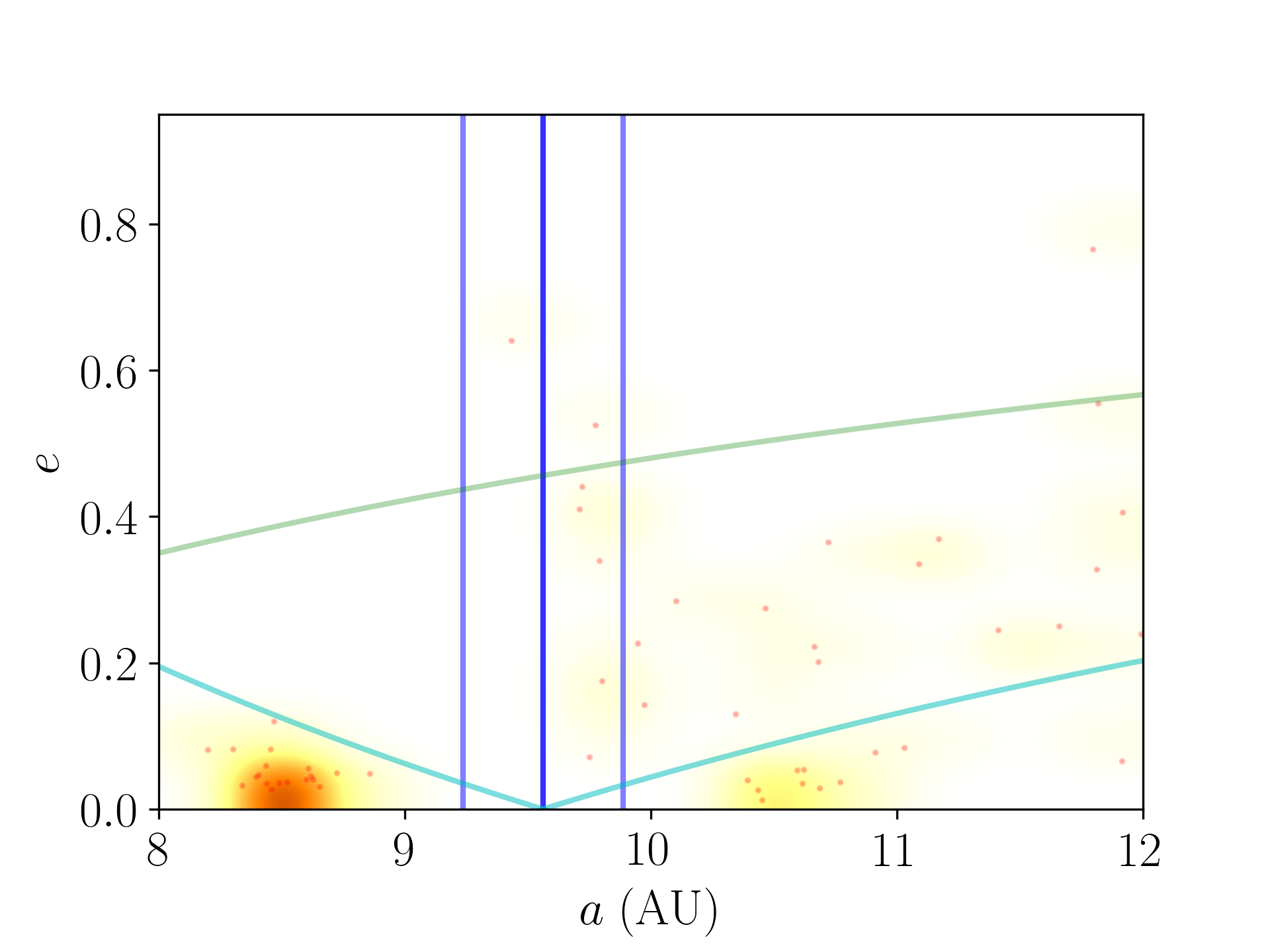}
        \caption{Near-planar case (mean $i = 175\degr$)}\label{fig:near_planar}
	\end{subfigure}
	\begin{subfigure}{0.33\textwidth}
        \centering
		\includegraphics[width=\textwidth]{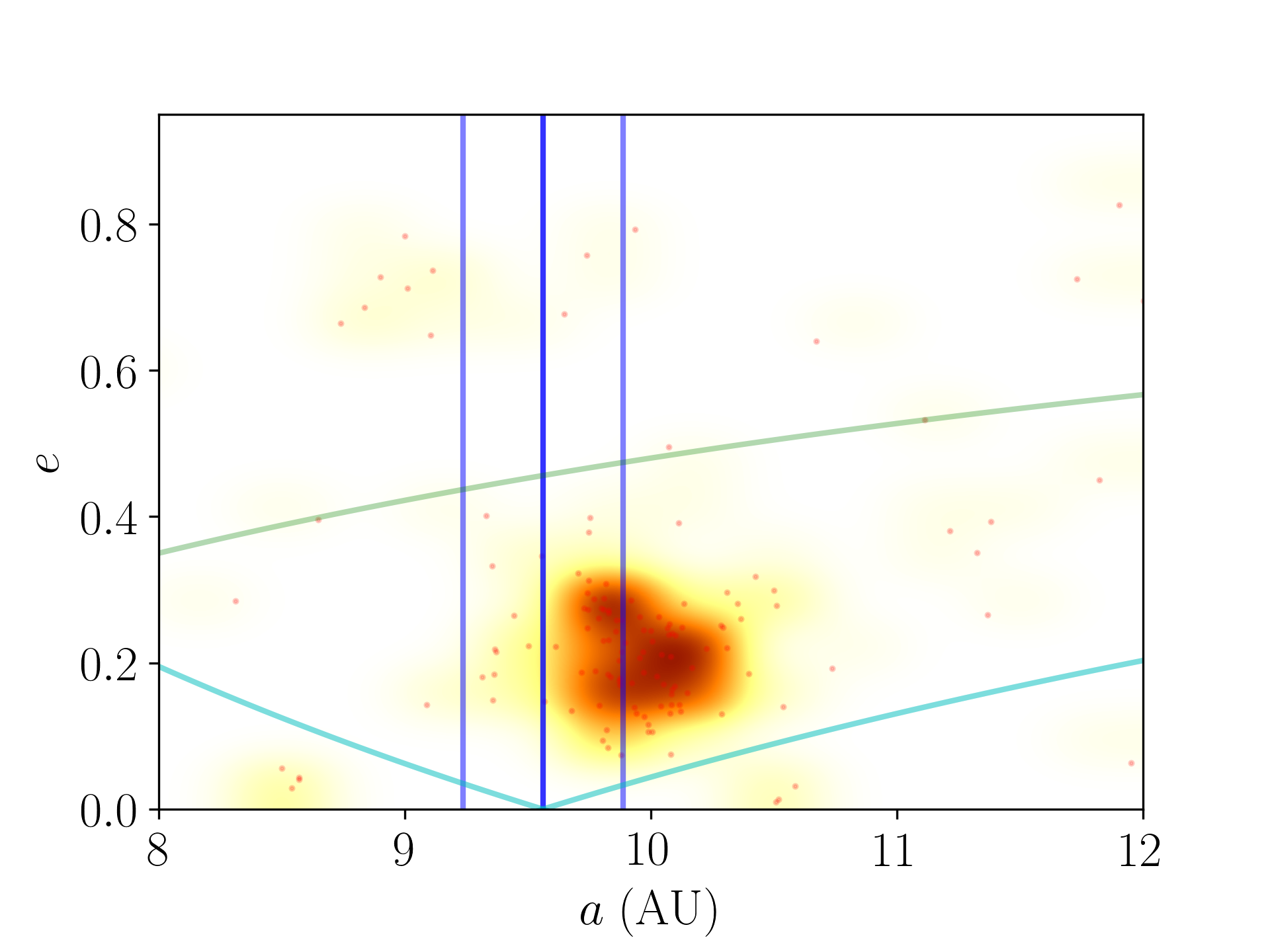}
        \caption{Inclined case (mean $i = 160\degr$)}\label{fig:inclined}
    \end{subfigure}
	\caption{Distributions and corresponding heat maps on the $(a,e)$ space of surviving particles in the planar case, the near-planar case and the inclined case are shown above. Every red dot denotes one remaining particle at the end of the $10$ Myr integration time. Heat maps are auto-generated with the relative density of dots to visualize the clusterings better. For each figure, three vertical blue lines indicate the centre of the retrograde 1:1 resonance with Saturn and its approximate borders. Two V-shaped curves are where the intersection at aphelion and perihelion of the orbit of a particle with those of Saturn (cyan) and Jupiter (green) occurs.}\label{fig:instability_ae}
\end{figure*}

\section{Numerical Surveys of Saturn's retrograde co-orbitals}\label{sec:2}
The most straightforward way to demonstrate the instability of Saturn's retrograde co-orbital region in the presence of Jupiter is to integrate a large number of test particles in different models. Such numerical surveys have been performed to prove that Jupiter is the culprit kicking Saturn Trojans out \citep{Hou:2013ic}. Therefore, we should check whether Jupiter is responsible for the instability of Saturn's retrograde co-orbitals.

We carried out three separate simulations considering all four giant planets (i.e. S-JSUN model) using the \textsc{mercury} \citep{1999MNRAS.304..793C} package. Each simulation contains over 10,000 test particles whose initial conditions are uniformly distributed on an $(a,e)$ grid, with $a$ ranging from 8.5 au to 10.5 au, and eccentricity $e$ ranging from 0 to 0.98. For the initial inclinations, three Rayleigh distributions of $\sin{i}$ with the mean value of $i = 180\degr$, $175 \degr$ and $160 \degr$ are used (or scale parameter $\sigma = 0$, $0.07$ and $0.27$, respectively). For the sake of simplicity, we call these three simulations the planar case, the near-planar case, and the inclined case. Additionally, the initial phase angles are picked randomly from $0$ to $2\pi$. Unlike the studies about Trojans, where phase angles were specifically selected to satisfy the $60\degr$ phase difference between the particle and the planet, we do not have to follow the $60\degr$ rule here, since asymmetric equilibrium points do not exist in the retrograde co-orbital resonance \citep{Huang:2018ey}.

For each simulation, we integrated these particles for $10$ Myr, with an output interval of $0.1$ Myr. The hybrid symplectic/Bulirsch-Stoer integrator was employed with a step size of $120$ days and an accuracy parameter of $10^{-12}$. The surviving particles and their corresponding heat maps are plotted in Figures~\ref{fig:instability_ae} to demonstrate the instability of the retrograde co-orbital region of Saturn in the S-JSUN model.

\subsection{Saturn's retrograde co-orbitals in the presence of Jupiter}\label{sec:2.1}

Here in Fig.~\ref{fig:instability_ae}, we present three scatter plots along with their heat maps of surviving particles at the end of the $10$ Myr integration time. Apparently, the retrograde co-orbital region of Saturn (marked by two light blue vertical lines) is highly unstable for a large range of eccentricities and inclinations. Moreover, we observed that test particles in this region would be normally cleared out within a timescale of $1$ Myr in three simulations.

For the planar and near-planar cases (left and middle panels of Fig.~\ref{fig:instability_ae}), the distributions of surviving particles are quite similar. Only particles with low eccentricities and below the intersection lines of Saturn (cyan curves in Fig.~\ref{fig:instability_ae}) have a chance to survive. Those with high eccentricities or originated from the co-orbital region eventually either collide with big planets or get scattered like Centaurs.
However, as shown in the right panel, the inclined case differs from the planar and near-planar cases. There is an obvious clustering of remaining particles around $a\sim10$ AU, $e\sim(0.1,0.3)$, and $i\sim(145\degr,165\degr)$, sitting at the right border of the co-orbital region. Although part of the clustering lies inside the resonant region, however, it does not imply that they are trapped in the retrograde 1:1 resonance with Saturn. Upon further investigation, we realize that they can stay there for 10 Myr due to Kozai-Lidov resonance with Saturn.

\begin{figure}
	\centering
	\includegraphics[width=\columnwidth]{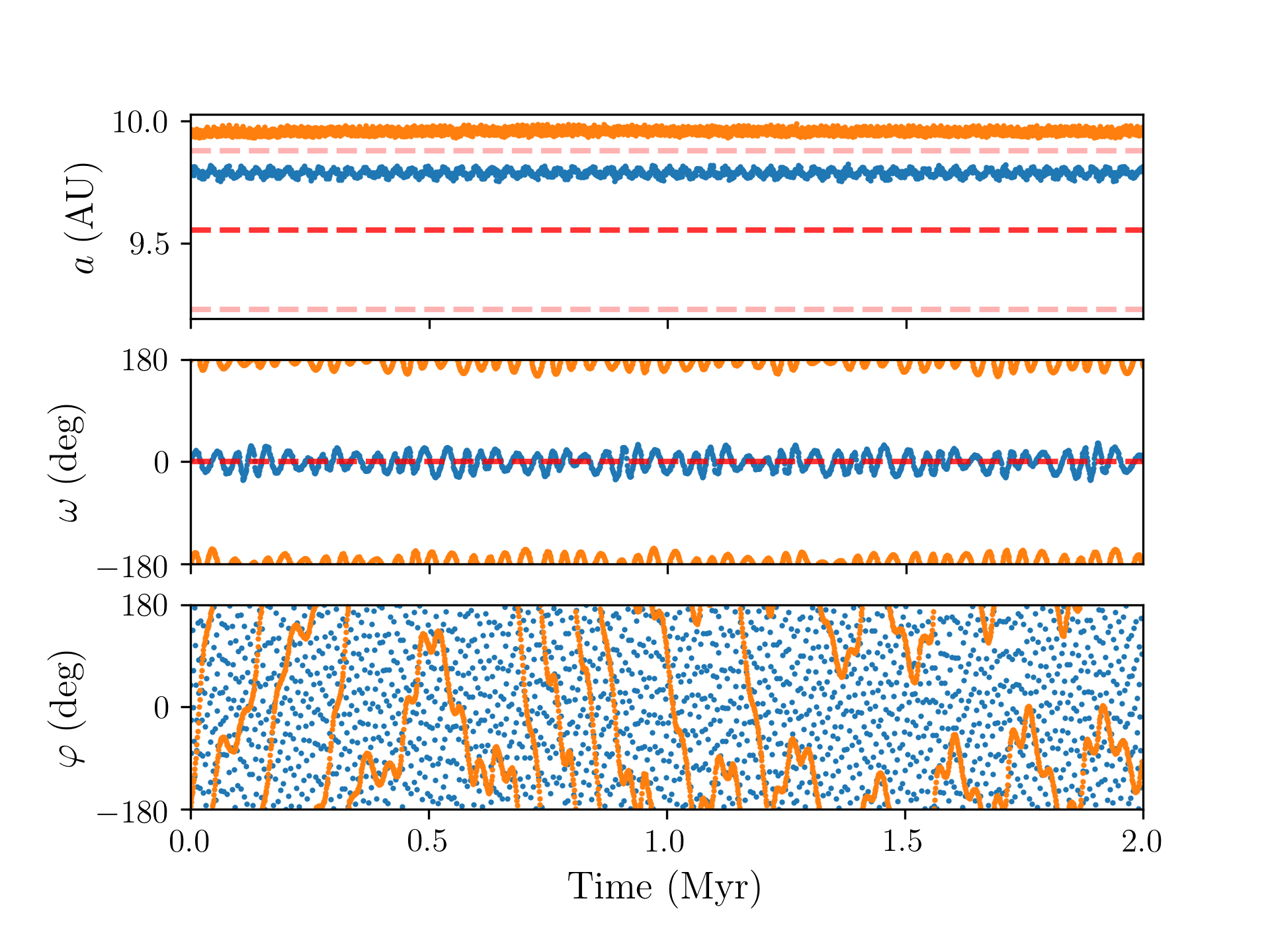}
	\caption{Dynamical evolution of two surviving particles from the clustering of our inclined case simulation (Fig.~\ref{fig:inclined}). The variations of their semi-major axes $a$, arguments of perihelion $\omega$, and resonant angles $\varphi$ are plotted in each panel. In the top panel, the retrograde 1:1 resonant centre of Saturn and its corresponding width are denoted by three red lines.}\label{fig:kozai_evo}
\end{figure}

Time evolutions of $a$, $\omega$ and $\varphi$ (i.e. the retrograde 1:1 resonant angle with Saturn) of two surviving particles in the inclined case are shown in Fig.~\ref{fig:kozai_evo}. The expression of the retrograde resonant angle is given by $\varphi = \lambda_{\textrm{par}} - \lambda_{\textrm{Saturn}} - 2\varpi_{\textrm{par}}$, where $\lambda_{\textrm{par}}$ and $\varpi_{\textrm{par}}$ are specifically defined for retrograde orbits \citep{2013CeMDA.117..405M}. As shown in the third panel of Fig.~\ref{fig:kozai_evo}, resonant angles of both test particles never librate, despite the fact that their semi-major axes always lies in the vicinity of the co-orbital region of Saturn (first panel of Fig.~\ref{fig:kozai_evo}). However, their arguments of perihelion $\omega$ do librate around $0\degr$ or $180\degr$, validating the Kozai-Lidov resonance state of both particles. Furthermore, We find that almost all of remaining particles in the clustering of Fig.~\ref{fig:inclined} must have their $\omega$ librating around either $0\degr$ or $180\degr$.

Unlike classical Kozai-Lidov cycles, where $\omega$ librates around $\pm 90\degr$ when $i$ exceeds a threshold \citep{Kozai:1962fa}, here the critical angle of Kozai-Lidov resonance librates around $0\degr$ or $180\degr$ because new librational regions would emerge if the semi-major axes of two bodies are very close \citep{Gronchi:wj}. This Kozai-Lidov libration provides a protection mechanism, which was was first proposed by \citet{Michel:1996us} to explain the stability of NEAs whose semi-major axes are close to that of Earth. The oscillation of $\omega$ around $0\degr$ or $180\degr$ protects an asteroid from close encounters with the perturbing planet, since the node crossings occur always near perihelion and aphelion. On condition that the orbit of the planet is almost circular and the eccentricity of the asteroid is high enough, its perihelion and aphelion are always far from the planet, meaning close encounters are not likely to happen, even though their orbits are indeed intersecting.

For retrograde bodies whose semi-major axes are close to that of a planet, the same protection mechanism applies. This is because the non-resonant Kozai-Lidov mechanism of a retrograde body should is consistent with its prograde counterpart, i.e., whose $I = 180\degr-I_\textrm{retro}$ \citep{Huang:2018cz}. One of the compelling examples appears in \citet{Namouni:2018hl} for the case of retrograde co-orbitals of Jupiter. It is demonstrated that all clones surviving the 4.5 billion years integration time around the co-orbital zone of Jupiter must be trapped inside the Kozai-Lidov resonance with $\omega = 0\degr$ or $180\degr$ while dynamically outside the retrograde 1:1 resonance. More interestingly, those long-term stable clones in their simulations also have semi-major axes above that of Jupiter, which is again consistent with our results in Fig.~\ref{fig:inclined}.

\subsection{Saturn's retrograde co-orbitals without Jupiter}\label{sec:2.2}

\begin{figure}
	\centering
	\includegraphics[width=\columnwidth]{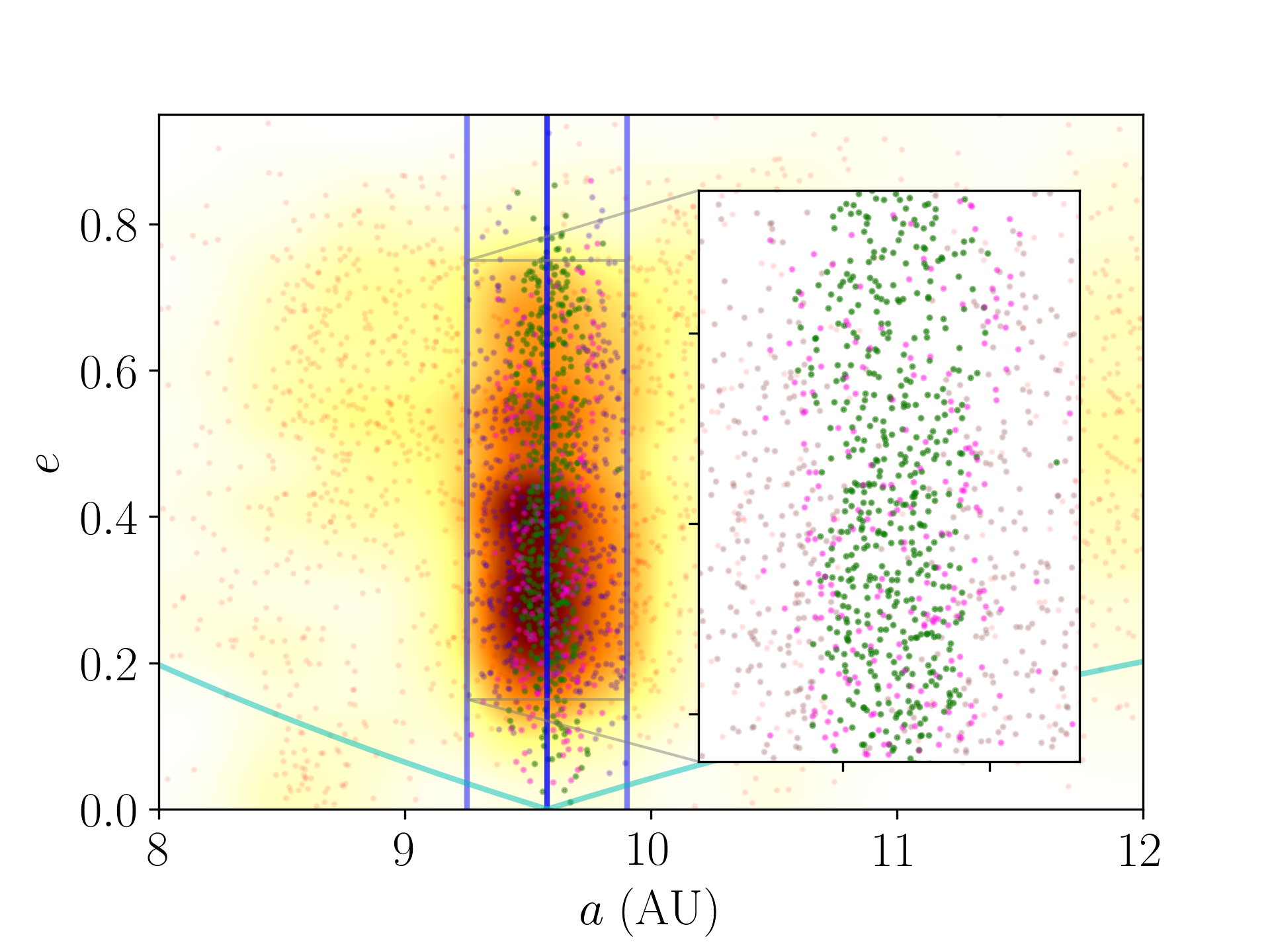}
	\caption{Distributions and the corresponding heat map on the $(a,e)$ space of surviving particles in the inclined case simulation without Jupiter. Green, grey, and pink dots inside Saturn's co-orbital space denote test particles stabilized by mean motion resonances, Kozai librations, and their compound effects, respectively. All other information is the same as Fig.~\ref{fig:instability_ae}.}\label{fig:inclined_ae_noJupiter}
\end{figure}

We have demonstrated above that the retrograde co-orbital region of Saturn is not stable in the S-JSUN model. However, we have not yet known whether the instability is caused by Jupiter, like the prograde case, or not. Therefore, we carried out another numerical simulation without considering the gravitational influence of Jupiter (i.e. S-SUN model). The results of this particular simulation are shown in Fig.~\ref{fig:inclined_ae_noJupiter}, with all initial conditions of test particles identical to the inclined case in Sec.~\ref{sec:2.1}.

As presented in Fig.~\ref{fig:inclined_ae_noJupiter}, most of the surviving particles lie between the 1:1 resonance borders, with eccentricities in a relatively large range of $e\sim(0.1,0.7)$. Upon careful inspection, we find that there are actually two mechanisms stabilizing the particles in the retrograde co-orbital region. The first one is the retrograde 1:1 resonance. 39.5\% of the co-orbital particles have their resonant angles $\varphi$ librating around $0\degr$, while $\omega$ circulating, showing a good resonant state with Saturn (denoted by green dots in Fig.~\ref{fig:inclined_ae_noJupiter}). And Kozai-Lidov resonances, again, provide another mechanism for the 30.7\% particles to survive (denoted by grey dots). Based on our last work \citep{Huang:2018cz}, the small-amplitude retrograde 1:1 resonance cannot coexist with the Kozai-Lidov resonance (shown in fig.2). In other words, the Kozai-Lidov libration only occurs when the resonant amplitude enlarges over $140\degr$ or the constraint of resonance is released. Therefore, we also observe that the compound effects of the large-amplitude resonance combined with the Kozai-Lidov libration stabilize 19.3\% particles (denoted by pink dots). The rest of 10.4\% are unstable particles temporarily inside the co-orbital region (denoted by red dots).

Unlike the inclined case in Fig.~\ref{fig:inclined}, there is not a single mechanism capable of creating such a huge co-orbital clustering in Fig.~\ref{fig:inclined_ae_noJupiter}. Statistically, a co-orbital closer to the precise resonant location is more likely to be affected by the mean motion resonance, while that near the borders is more likely to be controlled by the Kozai-Lidov mechanism. The compound effects serve as a transition between these two mechanisms, which is a dynamical feature well understood in \citet{Huang:2018cz}.

The discrepancy between simulation results obtained by S-JSUN model and S-SUN model implies that Jupiter plays a leading role in sweeping out potential retrograde co-orbitals of Saturn. Nevertheless, with these simulations, we cannot yet tell the exact mechanism bringing chaos to possible retrograde neighbours of Saturn.

\section{The instability of retrograde co-orbitals of Saturn}\label{sec:3}

With our prior knowledge on Saturn Trojans, it may lead us to conclude that the Great Inequality \citep{Lovett:1895kt}, or the 2:5 near-resonance between Jupiter and Saturn is the primary cause of the scenario. The overlap between the 2:5 outer resonance with Jupiter and 1:1 resonance with Saturn brings chaos to Saturn's co-orbital region and destabilizes potential Trojans \citep{Nesvorny:2002il}. Besides the mechanism of resonance overlapping, the secular resonance plays a significant role in shaping the dynamics of Saturn Trojans \citep{Marzari:2000eq,Hou:2013ic}. Therefore, these two mechanisms may also influence retrograde co-orbitals, which will be explored in this section.

\subsection{Great Inequality}\label{sec:3.1}

\begin{figure}
	\centering
	\includegraphics[width=\columnwidth]{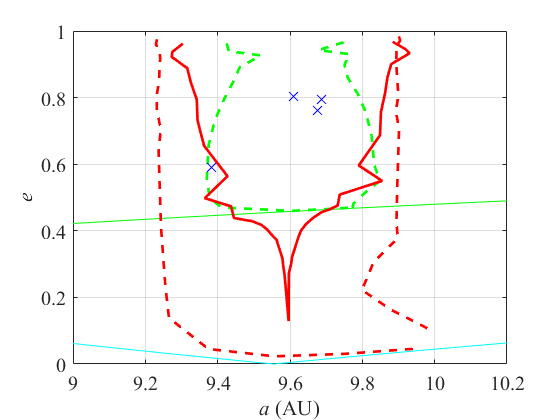}
	\caption{Resonance widths of retrograde 1:1 resonance with Saturn (outer red curve) and retrograde 2:5 resonance with Jupiter (inner red and green curves) on the $(a,e)$ space. Dashed curves denote librations around $0\degr$, while solid curve denotes that around $180\degr$. Orbit intersections with Jupiter and Saturn are shown by green and cyan curves. Four blue crosses denote locations of four retrograde minor bodies, 2006 RJ2, 2006 BZ8, 2012 YE8, and 2017 SV13, currently inside the co-orbital region of Saturn.}\label{fig:overlapping}
\end{figure}

To start with, it is worth noting that the prograde resonance and the retrograde resonance are significantly different. One of the major discrepancies is that they have a different order regarding the eccentricity. The order of a prograde $k:k^\prime$ resonance is defined as $|k-k^\prime|$. On the contrary, for a retrograde resonance of the same ratio, the order is $k+k^\prime$ instead \citep{2013CeMDA.117..405M}. The order of a resonance determines how its strength and width grow with its eccentricity. Generally speaking, a resonance with high order must have relatively low strength compared to a low order one, especially when its eccentricity is small. In our problem, a high order 7 for the retrograde 2:5 resonance hinders Jupiter from directly affecting the retrograde co-orbitals at a relatively low eccentricity, therefore, the retrograde resonance overlap is too weak to have a distinct impact on the dynamical evolution of retrograde co-orbitals.

In support of this argument, we measure widths of the retrograde 2:5 resonance with Jupiter and the retrograde 1:1 resonance with Saturn and plot them in Fig.~\ref{fig:overlapping}. Specifically, for eccentricities below the planet-crossing value, the width of the resonance is defined by the separatrix that bounds librations; For eccentricities above that value, the width of the resonance is defined by the trajectory of largest librational amplitude that does not cross the collision curve \citep{Morbidelli:2002tn}. With the workflow elucidated in \citet{Huang:2018ey}, we numerically generate a series of phase-space poitraits for both resonances and obtain their respective widths on the $(a,e)$ space. For the 2:5 resonance, both pericentric ($\varphi=0\degr$) and apocentric ($\varphi=180\degr$) libration centres are taken into account. However, for the 1:1 resonance, we only plot its width around $0\degr$ as its apocentric libration is weak enough to be ignored. It is worth mentioning that our resonance width of the retrograde 1:1 resonance is consistent with the numerical stability map by \citet[their fig. 6]{Morais:2016kq}.

As shown in Fig.~\ref{fig:overlapping}, librations around both $0\degr$ and $180\degr$ have relatively large widths when the eccentricity surpasses the threshold of 0.45, which is the crossing value of Jupiter. All minor bodies whose eccentricity exceeds this threshold will have their orbits intersecting with that of Jupiter and are therefore inherently unstable due to random close encounters, which is exactly the case of the four minor bodies currently inside Saturn's co-orbital region (denoted by crosses in Fig.~\ref{fig:overlapping}). As for potential retrograde co-orbitals with low eccentricity, the width of the 2:5 resonance around $180\degr$ is extremely narrow, compared to the size of the 1:1 resonance. Apparently, this is determined by its large order of 7. On the other hand, the resonance width of a prograde 2:5 resonance, which is of order 3, has a considerable size when the eccentricity is below 0.4 \citep{Malhotra:2018fh} and its overlapping effect is therefore non-negligible.

One of the tools to analyse the mean motion resonance overlap is the bi-circular model utilized by \citet{Nesvorny:2002il}, in which Jupiter and Saturn have a fixed motion in a planar circular orbit and their mutual interactions are ignored to rule out any secular factors. With such a simple model, it is shown that a V-shaped instability is inserted into the Saturn's co-orbital region, demonstrating that resonance overlap can raise chaos when $e>0.13$. Similarly, we also carry out another survey in the framework of bi-circular model to double check that retrograde resonance overlap cannot kick retrograde co-orbitals out.

\begin{figure}
	\centering
	\includegraphics[width=\columnwidth]{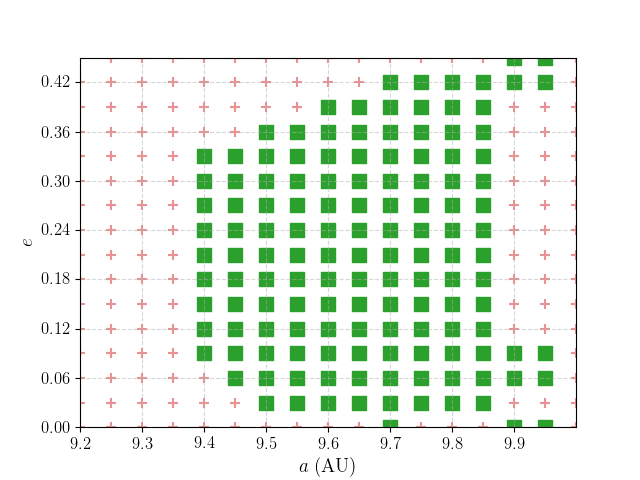}
	\caption{Stability Map of retrograde test particles in the bi-circular model. Every green square denotes a stable initial condition surviving for 10 Mry, while every red cross denotes an unstable one.}\label{fig:grid}
\end{figure}

As shown in Fig.~\ref{fig:grid}, another set of test particles, whose initial conditions are generated on a grid of $a\sim(9.2,10.0)$ and $e\sim(0,0.45)$, are integrated in the bi-circular model. After an integration time of 10 Myr, we plot particles always librating inside the co-orbital region as green squares, and those ejected as red crosses. In contrast to \citet{Nesvorny:2002il}, no instability caused by overlap is detected in Fig.~\ref{fig:grid}. Apparently, most of the stable particles lie between two retrograde 1:1 resonance borders shown in Fig.~\ref{fig:overlapping}, indicating their dynamics are predominantly shaped by co-orbital resonance with Saturn, rather than 2:5 resonance with Jupiter. With all of these associated pieces of evidence, therefore, we conclude that Great Inequality, i.e., the overlap between Jupiter's retrograde 2:5 resonance and Saturn's retrograde 1:1 resonance, is not a main cause for the instability of Saturn's retrograde co-orbitals.

% \footnote{Here, we choose an integration time of 0.1 Myr not 10 Myr because all secular terms have been ruled out in the bi-circular model. Therefore, the result will not alter even if the integration time is extended \citep{Nesvorny:2002il}.}

\subsection{Secular Resonances}\label{sec:3.2}

\begin{figure*}
	\centering
	\includegraphics[width=\textwidth]{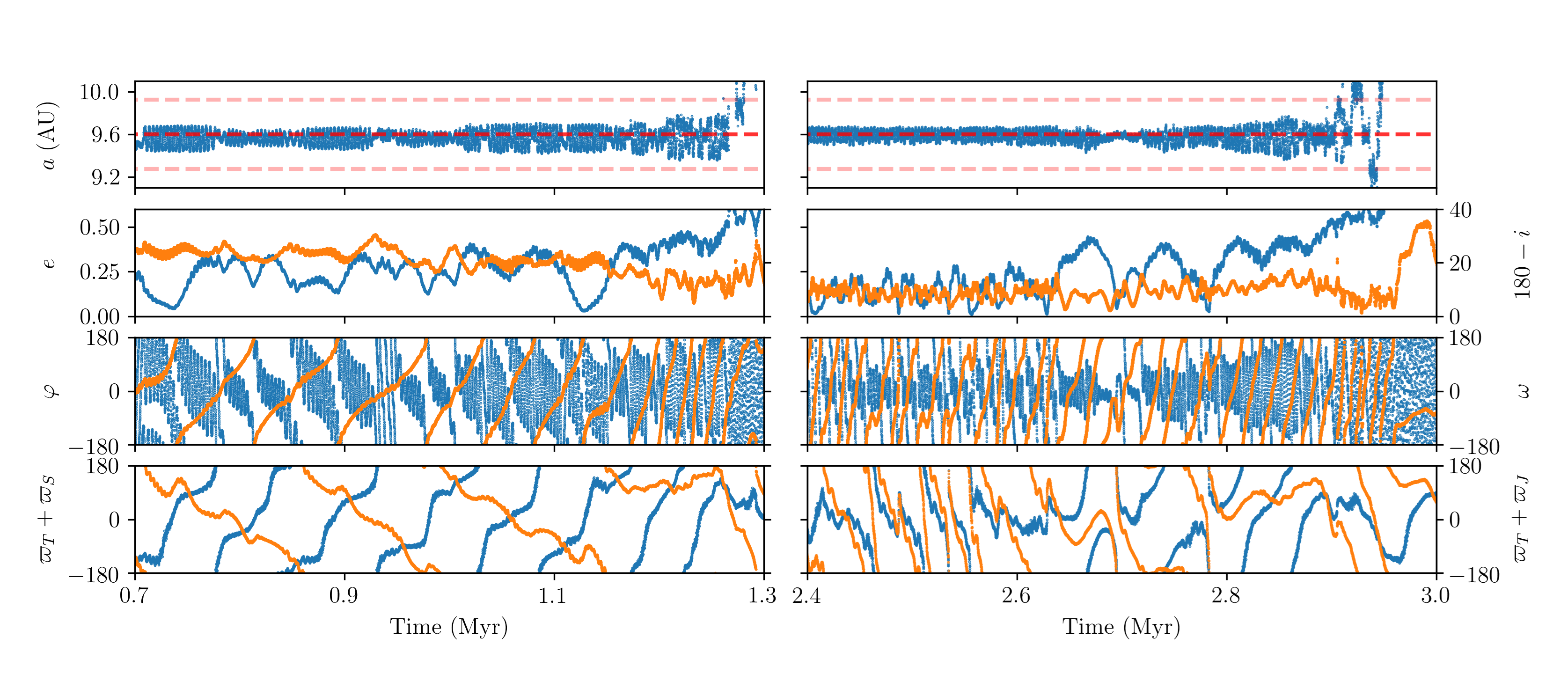}
	\caption{Orbital evolutions of two typical particles that get cleared out from Saturn's co-orbital space due to $\nu_5$ and $\nu_6$ secular resonances. Both particles are integrated in an S-JS model for a timescale of 4 Myr, but the left one has the initial inclination of $160\degr$, and the other has that of $180\degr$. In this plot, orbital elements on the left axis are denoted by blue curves, while those on the right axis are denoted by orange curves.}\label{fig:secular}
\end{figure*}

We have ruled out the possibility that the Great Inequality brings chaos to potential retrograde co-orbitals, as the strength of the 2:5 resonance with Jupiter is negligible when its eccentricity is below 0.4. Besides Great Inequality, secular resonances serve as another mechanism affecting Saturn's co-orbital region. In \citet{Marzari:2000eq}, it is proposed that the mixed $2\varpi_S - \varpi_J - \varpi_T$ resonance and possible $\nu_6$ resonance destabilize orbits of Saturn Trojans. As mentioned in \citet{Murray:2000id} and \citet{Morais:2001ej}, secular resonances occur when the proper precession frequency of the test body equals an eigen-frequency of the planetary system. In our case, the precession frequencies of the retrograde orbits are denoted by $-\dot{\varpi}_T$~\footnote{The definition of $\varpi_T$ is $\omega_T - \Omega_T$, which is actually the negative longitude of perihelion for retrograde orbits \citep{2013CeMDA.117..405M}. } and $\dot{\Omega}_T$. Therefore, the critical angles of the classical $\nu_5$,$\nu_6$, $\nu_{16}$, $2\nu_5-\nu_6$, and $2\nu_6-\nu_5$  secular resonances are now given by $\varpi_J + \varpi_T$, $\varpi_S + \varpi_T$, $\Omega_S - \Omega_T$, $2\varpi_J - \varpi_S + \varpi_T$, and $2\varpi_S - \varpi_J + \varpi_T$, respectively. However, it should be noted that these angles are just proxies that hint the presence of secular mechanisms, as they do not include any eigen-frequencies.

To validate the influence retrograde of secular resonances on co-orbitals, we further integrate 60 test particles in the S-JS model for 10 Myr and manually check all of their possible secular resonance states. All initial conditions have their semi-major axes equal to that of Saturn, eccentricities equal to 0.15, and phase angles randomnized. Three different inclinations, $180\degr, 170\degr$, and $160\degr$, are considered, 20 particles for each.

The initial eccentricity 0.15 is chose for the following reasons. Firstly, there is no need to inspect cases where $e$ is extremely low, because the left boundary of the retrograde 1:1 resonance crosses the low-eccentricity region horizontally (Fig.~\ref{fig:overlapping}), meaning a particle here will get strong disturbation from Saturn. The stability in this region is already very weak, which is supported by \citet[their fig. 7]{Morais:2016kq} and our Fig.~\ref{fig:inclined_ae_noJupiter}. Secondly, a nomincal route for a particle's escape from the co-orbital space would be gradually growing of the eccentricity till it surpasses the planet-crossing value of Jupiter ($e=0.45$). Therefore, as long as particles with a relatively low eccentricity, say 0.15, could be ejected from the co-orbital space, particles with higher eccentricity should get ejected, too.

Dynamical evolutions of two typical particles in this simulation, whose initial inclinations are $160\degr$ and $180\degr$, respectively, are presented in Fig.~\ref{fig:secular},  Unsurprisingly, as their eccentricities slowly grow up, almost all the particles get cleared out from Saturn's co-orbital space within a timescale of $4$ Myr. In addition, we find that, among secular resonances related to Jupiter and Saturn, $\nu_5$ and $\nu_6$ resonances play a vital role in raising particle's eccentricity to a higher value.

A $\nu_5$ or $\nu_6$ resonance crossing can be spotted from the change of the circulation direction of its corresponding critical argument \citep{Marzari:2000eq}. As shown in Fig.~\ref{fig:secular}, for the particle on the left, when its $\varpi_J + \varpi_T$ is slowly librating during the timespan from 1.1 Myr to 1.3 Myr, we notice its eccentricity undergoes a constant growth, surpassing the crossing value of Jupiter eventually. As for the particle on the right, we spot both $\nu_5$ and $\nu_6$ resonance crossings contributing to the gradual increase of the libration amplitudes of $a$ and $\varphi$. The combined impacts of two secular resonances boost the orbital eccentricity of the particle and undermine its resonant state with Saturn.

% And it is noteworthy that $\nu_5$ secular resonance is effective at any inclinations (left panel of Fig.~\ref{fig:secular}), but $\nu_6$ resonance crossing only happens for near-planar orbits whose inclinations are close to $180\degr$ (right panel of Fig.~\ref{fig:secular}).

% which occur twice around 0.2 Myr, once at 0.4 Myr, and once at the end time. Each time the particle crosses the secular resonance, its libration amplitudes of both $a$ and $\varphi$ will increase, and its $e$ will undergo a large jump, bring itself to an orbit closer to the territory of Jupiter. With its last resonance crossing at 0.55 Myr, the eccentricity of the body jumps above  to 0.5, ejected from Saturn's co-orbital region due to a random close encounter with Jupiter.

The dynamical evolution shown in Fig.~\ref{fig:secular} provides perfect examples showing how secular resonances boost the orbital eccentricity and finally destabilize a hypothetical retrograde co-orbital body. It is not a special case, but a general mechanism that we observed from almost all of these test particles. Statistically, 19/20 ($180\degr$), 16/20 ($170\degr$), 15/20 ($160\degr$) particles are destabilized by both $\nu_5$ and $\nu_6$ secular resonances; 1/20, 2/20, 3/20 particles are destabilized by $\nu_5$ resonance alone; 0/20, 2/20, 1/20 particles are destabilized by $\nu_6$ resonance alone. There is only one exception, whose intial inclination is $160\degr$, surviving the 10 Myr integration time. Unsurprisingly, this particle is well protected by the Kozai-Lidov libration around $0\degr$, as we expect.

For these particles, we also check other potential critical arguments, such as $2\varpi_J - \varpi_S + \varpi_T$, $2\varpi_S - \varpi_J + \varpi_T$, and $\Omega_S - \Omega_T$, but fail to find their correlations with instability as close as $\varpi_J + \varpi_T$ and $\varpi_S + \varpi_T$. Considering the simplicity of this model (Sun - Jupiter - Saturn - Retrograde co-orbitals), after ruling out factors like mean motion resonance overlap and other secular resonances, we believe it is safe to draw the conclusion that the instability is primarily due to the $\nu_5$ and $\nu_6$ resonances.

In conclusion, $\nu_5$ and $\nu_6$ resonances together can well explain the instability of the retrograde co-orbital region of Saturn and the nonexistence of potential retrograde co-orbitals in numerical simulations.

\section{Conclusion}\label{sec:4}
In this work, we find the absence of Saturn's hypothetical retrograde co-orbitals through numerical integrations. In the numerical surveys with the S-JSUN model, we observe an interesting clustering of test particles sitting near the right border of the co-orbital space, which is protected from close encounters with Saturn by Kozai-Lidov resonances around $0\degr$ and $180\degr$. In the numerical survey without Jupiter, we spot that both the mean motion resonance and the Kozai-Lidov libration are factors stabilizing Saturn's retrograde co-orbitals. Therefore, it is well demonstrated that the gravitational influence of Jupiter is responsible for the instability.

To explain precisely which mechanism provided by Jupiter shapes the co-orbital structure of Saturn, we examine two possible mechanisms, Great Inequality and secular resonances. On the one hand, we argue that the retrograde 2:5 resonance with Jupiter is too weak to alter the dynamics inside the co-orbital region of Saturn because its resonance width is narrow when eccentricity is below 0.4. Numerical simulations in the bi-circular model, in which all secular factors are ruled out, prove that the instability will not arise due to the retrograde mean motion resonance overlap.

On the other hand, through 60 test particles in the S-JS model, we manually check all of their critical angles corresponding to the secular resonances, and find secular resonances of $\nu_5$ and $\nu_6$ both conducive to the slow growth of orbital eccentricities, leading to the inevitable ejection with Jupiter of potential retrograde co-orbitals of Saturn.

To conclude, unlike the case of its prograde Trojans, the instability of Saturn's hypothetical retrograde co-orbitals is not induced by the Great Inequality, but seems to be primarily caused by the $\nu_5$ and $\nu_6$ secular resonances.

% \section{Semi-analytical Model}\label{sec:6}
% \subsection{Action-angle Variables for Retrograde MMRs}

% Following the process proposed by \citet{Morbidelli:2002tn}, we can deduce a set of canonical variables of the retrograde bi-circular problem:

% \begin{equation}\label{eq:retrograde_cononical_variables}
% 	\begin{aligned}
% 		\Sigma_1          & = \xi\left(\Lambda-\frac{k_2}{k_2+k''} P \right),\quad & \sigma_1          & = \frac{k_1 \lambda - k' \lambda'}{k_1+k'} + p,    \\
% 		\Sigma_2          & = -\xi\left(\Lambda-\frac{k_1}{k_1+k'} P \right),\quad & \sigma_2          & = \frac{k_2 \lambda - k'' \lambda''}{k_2+k''} + p, \\
% 		\tilde{\Lambda}'  & = \Lambda' + \frac{k'}{k_1+k'}\Sigma_1 ,\quad          & \tilde{\lambda}'  & = \lambda',                                        \\
% 		\tilde{\Lambda}'' & = \Lambda'' + \frac{k''}{k_2+k''}\Sigma_2 ,\quad       & \tilde{\lambda}'' & = \lambda'',                                       \\
% 	\end{aligned}
% \end{equation}
% where
% \begin{equation}\label{eq:xi}
% 	\xi = \frac{(k_1+k')(k_2+k'')}{k_1(k_2+k'')-k_2(k_1+k')}.
% \end{equation}

\section*{Acknowledgements}\label{sec:last}
We are very grateful to the anonymous referee for providing a thorough review of the paper, which led to a substantial improvement of the manuscript. This work is supported by the National Natural Science Foundation of China (Grant No.11772167).

%%%%%%%%%%%%%%%%%%%%%%%%%%%%%%%%%%%%%%%%%%%%%%%%%%



\bibliographystyle{mnras}
\bibliography{ref_3} % if your bibtex file is called example.bib

\begin{thebibliography}{}
\makeatletter
\relax
\def\mn@urlcharsother{\let\do\@makeother \do\$\do\&\do\#\do\^\do\_\do\%\do\~}
\def\mn@doi{\begingroup\mn@urlcharsother \@ifnextchar [ {\mn@doi@}
  {\mn@doi@[]}}
\def\mn@doi@[#1]#2{\def\@tempa{#1}\ifx\@tempa\@empty \href
  {http://dx.doi.org/#2} {doi:#2}\else \href {http://dx.doi.org/#2} {#1}\fi
  \endgroup}
\def\mn@eprint#1#2{\mn@eprint@#1:#2::\@nil}
\def\mn@eprint@arXiv#1{\href {http://arxiv.org/abs/#1} {{\tt arXiv:#1}}}
\def\mn@eprint@dblp#1{\href {http://dblp.uni-trier.de/rec/bibtex/#1.xml}
  {dblp:#1}}
\def\mn@eprint@#1:#2:#3:#4\@nil{\def\@tempa {#1}\def\@tempb {#2}\def\@tempc
  {#3}\ifx \@tempc \@empty \let \@tempc \@tempb \let \@tempb \@tempa \fi \ifx
  \@tempb \@empty \def\@tempb {arXiv}\fi \@ifundefined
  {mn@eprint@\@tempb}{\@tempb:\@tempc}{\expandafter \expandafter \csname
  mn@eprint@\@tempb\endcsname \expandafter{\@tempc}}}

\bibitem[\protect\citeauthoryear{Alexandersen, Gladman, Greenstreet, Kavelaars,
  Petit  \& Gwyn}{Alexandersen et~al.}{2013}]{Alexandersen:2013dq}
Alexandersen M.,  Gladman B.,  Greenstreet S.,  Kavelaars J.~J.,  Petit J.-M.,
   Gwyn S.,  2013, \mn@doi [Science] {10.1126/science.1238072}, 341, 994

\bibitem[\protect\citeauthoryear{Chambers}{Chambers}{1999}]{1999MNRAS.304..793C}
Chambers J.~E.,  1999, \mn@doi [MNRAS] {10.1046/j.1365-8711.1999.02379.x}, 304,
  793

\bibitem[\protect\citeauthoryear{Gronchi \& Milani}{Gronchi \&
  Milani}{1999}]{Gronchi:wj}
Gronchi G.~F.,  Milani A.,  1999, A\&A, 341, 928

\bibitem[\protect\citeauthoryear{Holman \& Wisdom}{Holman \&
  Wisdom}{1993}]{Holman:1993iw}
Holman M.~J.,  Wisdom J.,  1993, \mn@doi [AJ] {10.1086/116574}, 105, 1987

\bibitem[\protect\citeauthoryear{Hou, Scheeres  \& Liu}{Hou
  et~al.}{2013}]{Hou:2013ic}
Hou X.~Y.,  Scheeres D.~J.,   Liu L.,  2013, \mn@doi [MNRAS]
  {10.1093/mnras/stt1974}, 437, 1420

\bibitem[\protect\citeauthoryear{Huang, Li, Li  \& Gong}{Huang
  et~al.}{2018a}]{Huang:2018ey}
Huang Y.,  Li M.,  Li J.,   Gong S.,  2018a, \mn@doi [AJ]
  {10.3847/1538-3881/aac1bc}, 155, 262

\bibitem[\protect\citeauthoryear{Huang, Li, Li  \& Gong}{Huang
  et~al.}{2018b}]{Huang:2018cz}
Huang Y.,  Li M.,  Li J.,   Gong S.,  2018b, \mn@doi [MNRAS]
  {10.1093/mnras/sty2562}, 481, 5401

\bibitem[\protect\citeauthoryear{Innanen \& Mikkola}{Innanen \&
  Mikkola}{1989}]{Innanen:1989dv}
Innanen K.~A.,  Mikkola S.,  1989, \mn@doi [AJ] {10.1086/115036}, 97, 900

\bibitem[\protect\citeauthoryear{Kozai}{Kozai}{1962}]{Kozai:1962fa}
Kozai Y.,  1962, \mn@doi [AJ] {10.1086/108876}, 67, 579

\bibitem[\protect\citeauthoryear{Li, Huang  \& Gong}{Li
  et~al.}{2018}]{Li:2018kn}
Li M.,  Huang Y.,   Gong S.,  2018, \mn@doi [A\&A]
  {10.1051/0004-6361/201833019}

\bibitem[\protect\citeauthoryear{Lovett}{Lovett}{1895}]{Lovett:1895kt}
Lovett E.~O.,  1895, AJ, 15, 113

\bibitem[\protect\citeauthoryear{Malhotra, Lan, Volk  \& Wang}{Malhotra
  et~al.}{2018}]{Malhotra:2018fh}
Malhotra R.,  Lan L.,  Volk K.,   Wang X.,  2018, \mn@doi [AJ]
  {10.3847/1538-3881/aac9c3}, 156, 55

\bibitem[\protect\citeauthoryear{Marzari \& Scholl}{Marzari \&
  Scholl}{2000}]{Marzari:2000eq}
Marzari F.,  Scholl H.,  2000, \mn@doi [Icar] {10.1006/icar.2000.6376}, 146,
  232

\bibitem[\protect\citeauthoryear{Marzari, Tricarico  \& Scholl}{Marzari
  et~al.}{2003}]{Marzari:2003eh}
Marzari F.,  Tricarico P.,   Scholl H.,  2003, \mn@doi [A\&A]
  {10.1051/0004-6361:20031275}, 410, 725

\bibitem[\protect\citeauthoryear{Michel \& Thomas}{Michel \&
  Thomas}{1996}]{Michel:1996us}
Michel P.,  Thomas F.,  1996, A\&A, 307, 310

\bibitem[\protect\citeauthoryear{Morais}{Morais}{2001}]{Morais:2001ej}
Morais M. H.~M.,  2001, A\&A, 369, 677

\bibitem[\protect\citeauthoryear{Morais \& Namouni}{Morais \&
  Namouni}{2013}]{2013CeMDA.117..405M}
Morais M. H.~M.,  Namouni F.,  2013, \mn@doi [CeMDA]
  {10.1007/s10569-013-9519-2}, 117, 405

\bibitem[\protect\citeauthoryear{Morais \& Namouni}{Morais \&
  Namouni}{2016}]{Morais:2016kq}
Morais M. H.~M.,  Namouni F.,  2016, \mn@doi [CeMDA]
  {10.1007/s10569-016-9674-3}, 125, 91

\bibitem[\protect\citeauthoryear{Morbidelli}{Morbidelli}{2002}]{Morbidelli:2002tn}
Morbidelli A.,  2002, {Modern Celestial Mechanics: Aspects of Solar System
  Dynamics}.
Dynamics in the Solar System, CRC Press

\bibitem[\protect\citeauthoryear{Murray \& Dermott}{Murray \&
  Dermott}{1999}]{Murray:2000id}
Murray C.~D.,  Dermott S.~F.,  1999, {Solar System Dynamics}.
Cambridge University Press

\bibitem[\protect\citeauthoryear{Namouni \& Morais}{Namouni \&
  Morais}{2018}]{Namouni:2018hl}
Namouni F.,  Morais M. H.~M.,  2018, \mn@doi [MNRAS] {10.1093/mnrasl/sly057},
  477, L117

\bibitem[\protect\citeauthoryear{Nesvorn{\'{y}}}{Nesvorn{\'{y}}}{2002}]{Nesvorny:2002il}
Nesvorn{\'{y}} D.,  2002, \mn@doi [Icar] {10.1006/icar.2002.6961}, 160, 271

\bibitem[\protect\citeauthoryear{Wiegert, Connors  \& Veillet}{Wiegert
  et~al.}{2017}]{Wiegert:2017fj}
Wiegert P.,  Connors M.,   Veillet C.,  2017, \mn@doi [Nature]
  {10.1038/nature22029}, 543, 687

\bibitem[\protect\citeauthoryear{de~la Barre, Kaula  \& Varadi}{de~la Barre
  et~al.}{1996}]{delaBarre:1996he}
de~la Barre C.~M.,  Kaula W.~M.,   Varadi F.,  1996, \mn@doi [Icar]
  {10.1006/icar.1996.0073}, 121, 88

\makeatother
\end{thebibliography}

%%%%%%%%%%%%%%%%%%%%%%%%%%%%%%%%%%%%%%%%%%%%%%%%%%

% Don't change these lines
\bsp% typesetting comment
\label{lastpage}
\end{document}